\documentstyle[aps,prl,amssymb]{revtex}

\begin{document}
\draft
\title{SYMMETRY BREAKING of VIBRATING INTERFACES: A\ MECHANISM for MORPHOGENESIS\ }
\author{N.Garc\'{\i }a and V.V.Osipov}
\address{Laboratorio de F\'{\i }sica de Sistemas Peque\~{n}os y Nanotecnolog\'{\i }a,%
\\
Consejo Superior de Investigaciones Cient\'{\i }ficas, c/Serrano 144, 28006\\
Madrid, Spain}
\date{\today}
\maketitle

\begin{abstract}
We show that very small-amplitude oscillations of a highly symmetric,
spheric or cylindrical, interface (thin membrane) between two fluids can
result in inhomogeneous instability and breaking of the interface symmetry:
the frequency of the breathing vibration selects the spatial symmetry. This
mechanism may govern morphogenesis.
\end{abstract}


The nature of spontaneous symmetry breaking remains one of the most
enigmatic questions of modern science. This problem emerges in connection
with the equilibrium phase transitions, self-organization in nonequilibrium
systems and many other areas in physics, chemistry and biology (see, e.g., 
\cite{Prig}), as well as with cell fission and morphogenesis, i.e., the
development and spatial differentiation of complex body structures during
their growth \cite{Main}.

In 1952 Turing showed that the homogeneous state of some specific chemical
reactions can lose stability with regard to a spontaneous increase of
perturbations of certain form \cite{Tur}. Since then the chemical basis is
the prevalent idea of phenomenological theory of morphogenesis (see, e.g., 
\cite{Prig,Main}). Turing's model is based on chemical or biological
processes of local self-reproduction of some chemical agent (the activator)
and far-ranging inhibition. As a consequence of such processes a very small
increase of the activator concentration in a local region results in a
global redistribution of the substance concentrations and formation of more
complex structure \cite{Prig,Main}. However, the Turing's chemical reactions
are uncommon, unique and very complex processes.

In this work we develop a new mechanism, without complexity, that breaks the
symmetry by creating an instability in an oscillating interface, thin
membrane, separating two different fluids. In other words, we show that if,
for example, a spherical or cylindrical structure vibrates with a breathing
symmetric mode for a given set of the frequencies the symmetry breaks with
respect to bimodal, trimodal, pentagonal, etc. modes, i.e., the vibration
frequency selects the spatial symmetry of the interface.

We consider a thin symmetric membrane, spherical or cylindrical interface,
with the radius $R_{0}$ which separates two fluids with densities $\rho _{1}$%
and $\rho _{2}$ ($\rho _{1}\simeq $ $\rho _{2}\simeq \rho $) respectively.
Owing to Archimed's force the effective gravity acceleration operating on
the internal fluid is $g=g_{e}(1-\rho _{1}/\rho _{2})<<g_{e}.$ We propose $%
R_{0}$ is small enough, so the condition $\gamma k_{m}^{2}/\rho =\gamma
m^{2}/R_{0}^{2}\rho >>g$ is valid. Here $\gamma $ is the surface tension and 
$k=m/R_{0}$ is the typical wave vector of the increasing deformation of the
symmetric interface, $m=1,2,3,...$. This is the condition when we can
neglect the gravity and consider only the effect due to the surface tension
of the interface.

Let us take, at first, for definiteness, a spherically symmetric interface $%
{\bf S}$ whose radius, $R$, oscillates with a frequency $\omega :$ $%
R=R_{0}-d\cos \omega t.$ From the incompressibility of the fluid it follows
that its radial velocity is $v_{r0}=v_{R0}(t)R_{0}^{2}/r^{2},$ where $%
v_{R0}(t)=dR/dt=d\omega \sin \omega t$. (This means that some source, for
example, a small pulsating ball has to be inside the interface.) The
vortex-free motion of an ideal liquid (we consider the effect of the
viscosity below) is described by the Euler and the continuity equations:

\begin{equation}
\frac{d{\bf v}}{dt}=\frac{\partial {\bf v}}{\partial t}+({\bf v\nabla }){\bf %
v=-}\frac{1}{\rho }{\bf \nabla }p,  \label{Eul}
\end{equation}
\begin{equation}
{\bf \nabla }^{2}\Phi ={\bf \nabla }_{r}^{2}\Phi +{\bf \nabla }_{\perp
}^{2}\Phi =0  \label{cont}
\end{equation}
where $\Phi $ is the velocity potential, ${\bf v=\nabla }\Phi $ and ${\bf %
\nabla }_{\perp }^{2}$ is the part of Laplacian depending only on
coordinates of the surface ${\bf S}$. For the undistorted spherical surface,
from the symmetry of the problem, it follows that ${\bf v}_{\perp }=0,$
i.e., ${\bf \nabla }_{\perp }^{2}\Phi =0.$ Then, from Eq.\ref{cont}, we can
write that ${\bf \nabla }_{r}^{2}\Phi ={\bf \nabla }_{r}(v_{r0})=\partial
v_{r0}/\partial r+2v_{r0}/r=0,$ in accord with $%
v_{r0}=v_{R0}(t)R_{0}^{2}/r^{2}.$ In the presence of a distortion, $%
\varsigma $, of the spherical surface ${\bf S}$ the interface radial
velocity is $v_{r}=v_{R0}(t)+\partial \varsigma /\partial t.$ Using this, we
find from Eq.\ref{Eul} that near the interface 
\begin{equation}
\frac{dv_{r}}{dt}=F(t)+\partial ^{2}\varsigma /\partial t^{2}{\bf =-}\frac{1%
}{\rho }\frac{\partial p}{\partial r},  \label{vr}
\end{equation}

\begin{equation}
\frac{\partial {\bf v}_{\perp }}{\partial t}{\bf =-}\frac{1}{\rho }{\bf %
\nabla }_{\perp }p  \label{vx}
\end{equation}
where $F(t)=d\omega ^{2}\cos \omega t$ is the acceleration of the interface
and we neglect the term $({\bf v}_{\perp }{\bf \nabla }){\bf v}_{\perp }$ in
Eq. \ref{vx} ${\bf \ }$by virtue of smallness of $\varsigma $ \cite{LL}.
Owing to smallness of $\varsigma $ we can write the pressure near the
surface as

\begin{equation}
p=\rho F(t)(r-R-\varsigma )+\gamma {\bf \nabla }_{\perp }^{2}\varsigma
+p_{o}(t)  \label{pres}
\end{equation}
Here we took into account that the pressure at the interface (when $%
r=R+\varsigma )$ is $p=\gamma (\sigma _{1}+\sigma _{2})+p_{o}(t)$ where $%
\sigma _{1}$ and $\sigma _{2}$ are the principal curvatures of the interface 
\cite{LL}: $(\sigma _{1}+\sigma _{2})={\bf \nabla }_{\perp }^{2}\varsigma $
since ${\bf \nabla }_{\perp }^{2}\varsigma >R_{0}^{-1}$. Substituting Eq.\ref
{pres} into Eq.\ref{vx} we obtain 
\begin{equation}
\frac{\partial {\bf v}_{\perp }}{\partial t}=F(t){\bf \nabla }_{\perp
}\varsigma -\frac{\gamma }{\rho }{\bf \nabla }_{\perp }^{3}\varsigma \text{
or }\frac{\partial }{\partial t}{\bf \nabla }_{\perp }^{2}\Phi =F(t){\bf %
\nabla }_{\perp }^{2}\varsigma -\frac{\gamma }{\rho }{\bf \nabla }_{\perp
}^{4}\varsigma  \label{vy}
\end{equation}

We will seek solutions of the problem in the following form 
\begin{equation}
\zeta =\sum_{m=0}^{\infty }a_{m}(t)S_{m}\text{ and }\Phi =\sum_{m=0}^{\infty
}c_{m}(t)\Psi _{m}(r)S_{m}-v_{R0}(t)R_{0}^{2}/r  \label{t}
\end{equation}
where $S_{m}$ is the complete orthogonal set of eigenfunctions depending
only on the coordinates of the undisturbed surface ${\bf S}$ and satisfying
the following equation

\begin{equation}
({\bf \nabla }_{\perp }^{2}+k_{m}^{2})S_{m}=0  \label{Sm}
\end{equation}
for $r=R_{0}$ and the boundary conditions corresponding to the symmetry of
the problem. In the spherical case $S_{m}=C_{l,m}P_{l}^{\mid m\mid }(\cos
\theta )\exp (im\varphi )$ are the spherical functions of angles $\varphi $
and $\theta $ and $k_{m}^{2}=l(l+1)R_{0}^{-2}$ where $m=l,l-1,...,-l$ and $%
l=0,1,2,..$. Substituting $\Phi $ from Eq.\ref{t} into Eq.\ref{cont}, using
Eq.\ref{Sm} and the condition ${\bf \nabla }_{r}(v_{r0})=0$ cited above, we
obtain the equation for $\Psi _{m}(r):$

\begin{equation}
({\bf \nabla }_{r}^{2}-k_{m}^{2})\Psi _{m}(r)=0  \label{F}
\end{equation}
with the boundary conditions ${\bf \nabla }_{r}\Psi _{m}\rightarrow 0$ when $%
r\rightarrow 0$ and $\Psi _{m}(r)=A$ at $r=R_{0}$ where $A$ is some constant
which does not reveal itself in the final results. Near the interface ${\bf %
\nabla }_{r}\Phi =v_{r}=v_{R0}(t)+\partial \varsigma /\partial t$ and so
from Eq.\ref{t} it follows that $c_{m}(t)=da_{m}/dt({\bf \nabla }_{r}\Psi
_{m})_{r=R_{0}}^{-1}.$ Substituting $\Phi $ from Eq.\ref{t} into Eq.\ref
{cont} and using Eq.\ref{F} and $c_{m}(t),$ we find that

\begin{equation}
{\bf \nabla }_{\perp }^{2}\Phi =-\sum_{m=0}^{\infty }k_{m}^{2}\varkappa
_{m}^{-1}S_{m}da_{m}/dt  \label{dF}
\end{equation}
where $\varkappa _{m}=[{\bf \nabla }_{r}\Psi _{m}/\Psi _{m}(r)]_{r=R_{0}}$
does not depend on the constant $A$ . Then from Eq.\ref{dF} and Eq.\ref{vy},
we obtain

\begin{equation}
d^{2}a_{m}/dt^{2}+[\gamma k_{m}^{2}\varkappa _{m}\rho ^{-1}-\varkappa
_{m}F(t)]a_{m}=0.  \label{ATT}
\end{equation}
Using $T=\omega t/2$ we can rewrite Eq.\ref{ATT} as 
\begin{equation}
d^{2}a_{m}/dT^{2}+(p_{m}-2q_{m}\cos \omega t)a_{m}=0,  \label{at}
\end{equation}
where

\begin{equation}
q_{m}=2\varkappa _{m}d\text{ and }p_{m}=\Omega _{m}^{2}\omega ^{-2}\text{
where }\Omega _{m}^{2}=4k_{m}^{2}\varkappa _{m}\gamma \rho ^{-1}.  \label{pq}
\end{equation}
For the spherical interface $k_{m}R_{0}>1$ and $\varkappa _{m}\simeq k_{m}=$ 
$[l(l+1)]^{1/2}R_{0}^{-1}$ and so $\Omega
_{m}^{2}=4[l(l+1)]^{3/2}R_{0}^{-3}\gamma \rho ^{-1}$ and $%
q_{m}=2d[l(l+1)]^{1/2}R_{0}^{-1}.$

These results can be extended easily to other cases. For example, when the
interface have a form of a cylinder with vibrating radius, then $S_{m}=\cos
(k_{l}z)\exp (im\varphi )$ and in Eq.\ref{pq} $\varkappa _{m}\simeq k_{m}$
and $k_{m}^{2}=m^{2}/R_{0}^{2}+\pi ^{2}l^{2}/h_{0}^{2}$ where $h_{0}$ is
height of the cylinder. This vibrating cylindrical body can spontaneously
distort in the axis $z$ or with respect the azimuthal perturbations.

We emphasize that Eq.\ref{at} coincides with Eq.(2.12) of Ref. \cite{Benj}
to describe the Faraday's instability \cite{Far} of the plane free surface
of an ideal liquid under vertical periodic vibrations. These equations
differ in the values of the parameters $p_{m}$ and $q_{m}$. Moreover, in
contrast to the Faraday's instability when the vibrations are reduced to
trivial renormalization of the gravity, in this work we consider spherical
or cylindrical oscillating interfaces when the vertical direction, axial
gravity, is not distinguished from other directions. Benjamin and Ursell 
\cite{Benj} have constructed the stability diagram for Eq.\ref{at} with
respect to the universal parameters $p_{m}$ and $q_{m}$ using the analogy
between Eq.\ref{at} and the Mathieu's equation \cite{Math}. From this
diagram it follows that the instability is realized only in regions near the
points $p_{m}=n^{2}$ where $n=1,2,3,4,...$. In other words, the condition

\begin{equation}
\omega =\omega _{n,m}\simeq n^{-1}\Omega _{m}=2n^{-1}k_{m}(\varkappa
_{m}\gamma /\rho )^{1/2}  \label{pn}
\end{equation}
determines the resonant vibration frequencies when the symmetric interface
spontaneously deforms with respect to the standing wave with the azimuthal
number $m$. However, the greater is $n$, the narrower is the width $%
E_{g}^{(n)}(q_{m})$ of the $n$-th region of the instability for given $q_{m}$
\cite{Benj,Math}. For the widest instability region, with $n=1,$ the value $%
E_{g}^{(1)}(q_{m})\simeq 2q_{m}$ for $q_{m}<1$. It means that the
instability takes place when $(1-q_{m})<p_{m}<(1+q_{m}),$ i.e., the symmetry
breaking is realized for the vibration frequency lying within the spectral
range:

\begin{equation}
\Omega _{m}(1-\varkappa _{m}d)<\omega <\Omega _{m}(1-\varkappa _{m}d).
\label{omega}
\end{equation}

The threshold of the vibration amplitude $d$ is limited by the fluid
viscosity. For real fluid Eqs.\ref{ATT} and \ref{at} include the additional
terms $\gamma _{m}da_{m}(t)/dt$ and $\Gamma _{m}da_{m}(T)/dT$, respectively,
where $\gamma _{m}=2\nu k_{m}^{2}C_{1m}$ and $\Gamma _{m}=4\nu
k_{m}^{2}C_{1m}/\omega $ are proportional to the kinematic viscosity $\nu $
and $C_{m}$ is some constant of the order of unity \cite{Cerda}. The
threshold vibration amplitude, $d=d_{t},$ for the instability region with $%
n=1,$ can be estimated from the condition {\it \ }$E_{g}^{(1)}(q_{m})>2%
\Gamma _{m}$, i.e., $q_{m}=2\varkappa _{m}d>\Gamma _{m}.$ This condition
follows practically from results of Refs.\cite{Cerda} and is obtained in 
\cite{GO}. Using Eq.\ref{omega}, this condition can be written as 
\begin{equation}
d>d_{t}=2\nu C_{m}k_{m}^{2}/\varkappa _{m}\omega \simeq \nu (\rho /\gamma
\varkappa _{m})\simeq \nu (\rho R_{0}/\gamma m).  \label{dt}
\end{equation}
For parameters of water $d_{t}\simeq 4\mu m$, i.e., the threshold vibration
is a very small flutter of the interface.

We propose that the results above may be used as a basis for a simple,
without complexity, mechanism to trigger the fanciful morphogenesis
appearing in nature. The frequency of homogeneous interface vibrations
self-selects the space symmetry. If the interface oscillates with a
characteristic frequency the germ symmetry will break when its radius $R_{0}$
amounts to the quantity satisfied by Eq.\ref{omega} or Eq.\ref{pn} for $n>1$%
. After the new symmetry appears the growth rate increases with surface
curvature as is usual for many of Stefan-like problems \cite{Grow}.

The mathematical results reported here will be applied in a forthcoming
paper to explain the morphogesis of the acetabularia, equinoderms and cell
fision.

This work has been supported by the Spanish DGCIYT and by a NATO fellowship
Grant.


\end{document}